\documentclass[12pt]{iopart}

\usepackage{graphicx}
\usepackage{epstopdf}
\usepackage{makecell}
\usepackage{array}
\usepackage{multirow}
\usepackage{gensymb}
\newcolumntype{x}[1]{>{\centering\hspace{0pt}}p{#1}}
\usepackage{iopams}
\usepackage{longtable}
\usepackage{xcolor}

\begin{document}

\title[Superconductivity-induced phonon spectra renormalization]{Evidence of superconductivity-induced phonon spectra renormalization in alkali-doped iron selenides}

\author{M Opa\v{c}i\'c$^1$, N Lazarevi\'c$^1$, M \v{S}\'cepanovi\'c$^1$, Hyejin Ryu$^{2,3}$, Hechang Lei$^2$, C Petrovic$^{2,3}$ and Z V Popovi\'c$^1$}
\address{$^1$ Center for Solid State Physics and New Materials, Institute of Physics Belgrade, University of Belgrade, Pregrevica 118, 11080 Belgrade, Serbia}
\address{$^2$ Condensed Matter Physics and Materials Science Department, Brookhaven National Laboratory, Upton, New York 11973-5000, USA}
\address{$^3$ Department of Physics and Astronomy, Stony Brook University, Stony Brook, New York 11794-3800, USA}
\ead{nenadl@ipb.ac.rs}

\date{\today}

\begin{abstract}
Polarized Raman scattering spectra of superconducting K$_x$Fe$_{2-y}$Se$_2$ and nonsuperconducting K$_{0.8}$Fe$_{1.8}$Co$_{0.2}$Se$_2$ single crystals were measured in a temperature range from 10 K up to 300 K. Two Raman active modes from the $I4/mmm$ phase and seven from the $I4/m$ phase are observed in frequency range from 150 to 325 cm$^{-1}$ in both compounds, suggesting that K$_{0.8}$Fe$_{1.8}$Co$_{0.2}$Se$_2$ single crystal also has two-phase nature. Temperature dependence of Raman mode energy is analyzed in terms of lattice thermal expansion and phonon-phonon interaction. Temperature dependence of Raman mode linewidth is dominated by temperature-induced anharmonic effects. It is shown that change of Raman mode energy with temperature is dominantly driven by thermal expansion of the crystal lattice. Abrupt change of the A$_{1g}$ mode energy near $T_C$ was observed in K$_x$Fe$_{2-y}$Se$_2$, whereas it is absent in non-superconducting K$_{0.8}$Fe$_{1.8}$Co$_{0.2}$Se$_2$. Phonon energy hardening at low temperatures in the superconducting sample is a consequence of superconductivity-induced redistribution of the electronic states below critical temperature.
\end{abstract}

\maketitle

\section{Introduction}

Since the discovery of superconductivity in FeSe layered compound K$_x$Fe$_{2-y}$Se$_2$ \cite{Superconductivity}, considerable attention was focused on iron selenide materials due to their relatively high superconducting transition temperatures ($T_C$).\cite{Stjuart, LiuLuo,  Supercondmag, Litvinchuk} Recent investigations of single-layer FeSe films grown on SrTiO$_3$ revealed superconductivity with $T_C$ above 100 K, which is the highest critical temperature among all iron based materials discovered so far.\cite{Priroda2} Alkali metal-doped iron chalcogenides have some interesting features that distinguish them from the other iron based superconductors.\cite{Dagotto, Bao} Angle resolved photoemission measurements showed that there are no hole pockets at the Fermi level in K$_x$Fe$_{2-y}$Se$_2$, which opens the possibility of different type of pairing mechanism than in the iron pnictides. \cite{Gap} This is consistent with observed negative values of the Hall constant $R_H$ in a temperature range 0 - 150 K \cite{Hall}, which implies that conduction is predominantly governed by electron-like carriers.

The most striking feature of K$_x$Fe$_{2-y}$Se$_2$ single crystals is the presence of two distinct phases: insulating and metallic/superconducting. \cite{Priroda, Nanoscale, Bao} Insulating phase has antiferromagnetically, $\sqrt{5} \times \sqrt{5}$, ordered Fe vacancies with large iron magnetic moments, whereas superconducting phase is free of vacancies.\cite{Priroda} Resistivity measurements on the sample with nominal composition K$_{0.8}$Fe$_2$Se$_2$ revealed that superconductivity occurs below $T_C$$\sim$30 K.\cite{Superconductivity} However, it was later established that superconductivity appears only in Fe-deficient samples. Broad hump in the in-plane resistivity $\rho_{ab}(T)$, whose position varies between 105-240 K (depending on the sample preparation), pressumably occurs due to the type of connection between two phases.\cite{Hall, Precipitate} Below the hump, K$_x$Fe$_{2-y}$Se$_2$ is metallic due to an intrinsic property of metallic/superconducting state, since insulating regions do not contribute to the spectral weight close to Fermi energy, as observed by Angle Resolved Photoemission Spectroscopy.\cite{Riggs, Influence} It was also shown that M-doping on the Fe-site (M=Cr, Co, Ni, Zn) strongly suppresses superconductivity.\cite{Doping, Serija} However, the correlation between antiferromagnetically ordered iron vacancies and superconductivity remains unclear.

Raman scattering study of K$_x$Fe$_{2-y}$Se$_2$ single crystals has been performed by several groups.\cite{Zhang, Lazarevic1, Blumberg}. Zhang \textit{et al.} \cite{Zhang} considered phonon properties of this compound in terms of $I4/m$ space group. They observed and assigned 11 out of 18 Raman active modes. Their analysis revealed that phonon of unknown origin, at about 180 cm$^{-1}$ (which was assigned as A$_g$ symmetry mode) exhibits abrupt hardening of $\sim$1 cm$^{-1}$ at the superconducting critical temperature $T_C$. Lazarevi\'c \textit{et al.} \cite{Lazarevic1} analyzed lattice dynamics of K$_x$Fe$_{2-y}$Se$_2$ in terms of $I4/m$ and $I4/mmm$ phase. They observed and assigned two (of two possible) modes from the high-symmetry phase and 16 out of 18 phonon modes from the low-symmetry phase. They also observed Raman mode at about 180 cm$^{-1}$ and assigned them as of the A$_{1g}$ symmetry originating from superconducting I4/mmm phase. In \cite{Blumberg}, authors argued that new Raman modes, at about 165, 201 and 211 cm$^{-1}$, appear with cooling the sample below 250 K. Based on this, it is concluded that K$_x$Fe$_{2-y}$Se$_2$ single crystal exhibits a structural phase transition from $I4/m$ to $I4$ space group. They assumed a Fano-like shape for some phonons and analyzed temperature dependence of Raman mode energy and linewidth in terms of lattice anharmonicity. To the best of our knowledge, Raman spectrum of the K$_x$Fe$_{2-y}$Se$_2$ single crystals doped with Co is unknown.

In order to determine the impact of superconductivity on phonon properties of alkali-doped iron selenides, we have performed Raman scattering study of superconducting K$_x$Fe$_{2-y}$Se$_2$ and nonsuperconducting K$_{0.8}$Fe$_{1.8}$Co$_{0.2}$Se$_2$ single crystals. Raman spectra were measured in a temperature range from 10 to 300 K. All phonons expected to appear in the investigated energy range, according to \cite{Lazarevic1}, are observed in both single crystals. A detailed analysis of energy and linewidth temperature dependence for seven observed Raman active modes is performed.

\begin{figure}
\begin{center}
\includegraphics[width = 0.45\textwidth]{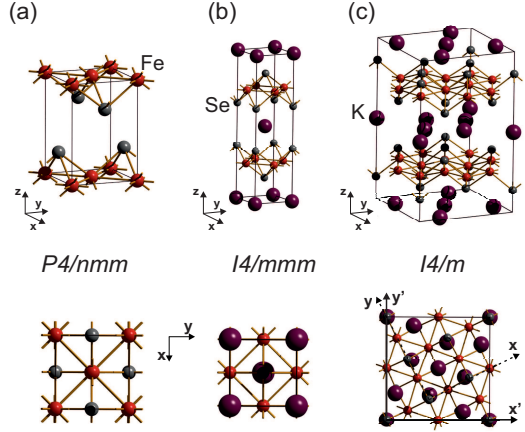}
\caption{(Color online) Unit cell for (a) FeSe, (b) and (c) K$_x$Fe$_{2-y}$Se$_2$  single crystals. In the right lower part I4/m and I4/mmm unit cells are shown together ($\mathbf{x,y}$ denotes crystallographic axes of the I4/mmm phase, whereas $\mathbf{x', y'}$ are crystallographic axes of the I4/m phase domain).}
\label{fig1}
\end{center}
\end{figure}

\section{Experiment}

K$_{x}$Fe$_{2-y}$Se$_2$ and K$_{0.8}$Fe$_{1.8}$Co$_{0.2}$Se$_2$ single crystals were grown and characterized as described elsewhere in details.\cite{Lei} Raman scattering measurements were performed on (001)-oriented samples, using a TriVista 557 Raman system, in backscattering micro-Raman configuration. The 514.5 nm line of an Ar$^+/$Kr$^+$ mixed gas laser was used as an excitation source. The Raman scattering measurements were carried out at low laser power, in order to minimize local heating of the sample. All measurements were performed in the vacuum, using KONTI CryoVac continuous flow cryostat with 0.5 mm thick window. The samples were cleaved just before the placement in the cryostat in order to obtain flat shiny surface. For the data extraction from the Raman spectra, Voigt profile has been used, where Gaussian width of 2 cm$^{-1}$ represents spectral resolution of the instrument.

FeSe layer is the basic building block for all iron selenide superconductors.  FeSe single crystal (see Figure~\ref{fig1}(a)) crystallizes in the tetragonal crystal lattice ($P4/nmm$ space group), with two formula units per unit cell. \cite{Margadonna} Site symmetries of individual atoms in this space group are $D_{2d}$ (Fe) and $C_{4v}$ (Se). Factor group analysis gives the normal mode distribution in the Brillouin zone center:
\begin{eqnarray}
(D_{2d}): \Gamma = A_{2u}+B_{1g}+E_g+E_u, \nonumber \\
(C_{4v}): \Gamma = A_{1g} + A_{2u} + E_g + E_u. \nonumber
\end{eqnarray}
\noindent In total,  A$_{1g}(\alpha_{xx+yy}, \alpha_{zz})$, B$_{1g}(\alpha_{xx-yy})$ and 2E$_g(\alpha_{xz}, \alpha_{yz})$ phonons are Raman active. A$_{1g}$ (B$_{1g}$) modes represent vibrations of Se (Fe) ions along the $z$-axis, whereas $E_g$ modes originate from vibrations of Fe and Se ions within the $ab$ plane. \cite{Gnezdilov1}

\begin{figure}
\begin{center}
\includegraphics[width = 0.45\textwidth]{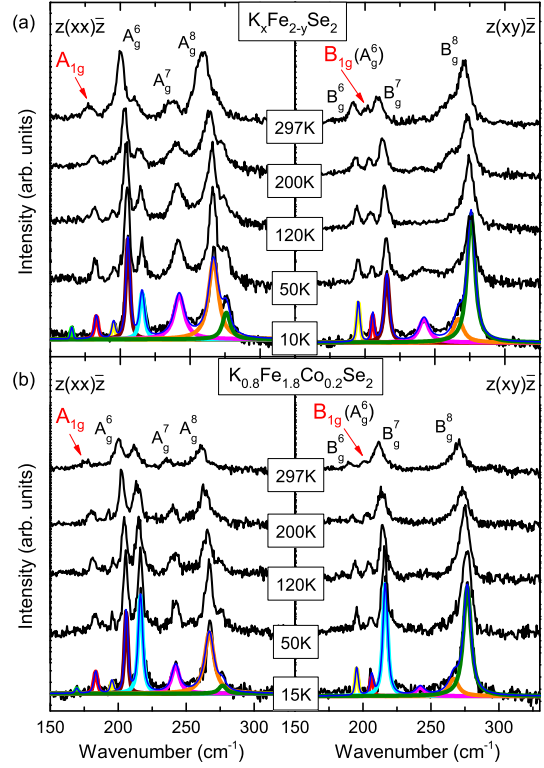}
\caption{(Color online) Temperature dependent Raman spectra of (a) K$_{x}$Fe$_{2-y}$Se$_2$ and (b) K$_{0.8}$Fe$_{1.8}$Co$_{0.2}$Se$_2$ single crystals in parallel (left panel) and crossed (right panel) polarization configuration ($\mathbf{x}=[100], \mathbf{y}=[010]$). Phonon modes originating from the high symmetry $I4/mmm$ phase are marked by arrows.}
\label{fig2}
\end{center}
\end{figure}

K$_x$Fe$_{2-y}$Se$_2$ single crystal consists of K ions intercalated between the FeSe slabs, which dominantly determines its physical properties.\cite{Superconductivity} At high temperature it crystallizes in the $I4/mmm$ space group. With cooling below 532 K partial symmetry breaking occurs, as a consequence of Fe vacancy ordering.\cite{Bao} Recent studies revealed that, at low temperatures, K$_x$Fe$_{2-y}$Se$_2$ single crystals consist of two phases separated at the nanometer scale: superconducting phase ($I4/mmm$) and insulating phase ($I4/m$).\cite{Priroda, Nanoscale, Lazarevic1} By comparing the crystallographic data for FeSe \cite{Margadonna} and K$_x$Fe$_{2-y}$Se$_2$ \cite{Superconductivity}, it can be seen that the intralayer Fe-Fe distances and Fe-Se bond lengths increase in K$_x$Fe$_{2-y}$Se$_2$ only by few percents.

Site symmetries of atoms in the $I4/mmm$ group (see Figure~\ref{fig1}(b)) are $D_{4h}$ (K), $C_{4v}$ (Se) and $D_{2d}$ (Fe). By applying the symmetry analysis, it follows:
\begin{eqnarray}
(D_{4h}): \Gamma = A_{2u}+E_u, \nonumber \\
(D_{2d}): \Gamma = A_{2u}+B_{1g}+E_g+E_u, \nonumber \\
(C_{4v}): \Gamma = A_{1g}+A_{2u}+E_g+E_u, \nonumber
\end{eqnarray}
\noindent giving the same Raman active modes distribution in the Brillouin zone center as in the case of FeSe. When measuring in the $(001)$-plane of the sample, only A$_{1g}$ and B$_{1g}$ modes are observable in the Raman scattering experiment.\cite{Nikl} In \cite{Lazarevic1}, these modes have been observed at about 180 and 207 cm$^{-1}$ (at 85 K), respectively. Gnezdilov \textit{et al.} \cite{Gnezdilov1} recently showed that A$_{1g}$ and B$_{1g}$ phonons in FeSe single crystal, originating from the same vibrations as in the case of K$_x$Fe$_{2-y}$Se$_2$, appear in the Raman scattering spectra at similar frequencies ($\sim$182 and $\sim$205 cm$^{-1}$, respectively, at 80 K). Together with \textit{ab initio} phonon calculations for the $I4/mmm$  space group with magnetic ordering included, \cite{Coupling} this confirms the assignation given in \cite{Lazarevic1}.

$I4/m$ crystal structure is shown in Figure~\ref{fig1}(c). Lower part of Figure~\ref{fig1}(c) illustrates I4/m and I4/mmm unit cells projected on the (001)-plane of the sample. It should be noted that crystallographic axes of the I4/m phase ($\mathbf{x', y'}$) are rotated by an angle $\alpha \approx 26.6^0$ with respect to the axes ($\mathbf{x,y}$) of the I4/mmm phase, due to $\sqrt{5}\times \sqrt{5}$ vacancy order. Site symmetries of atoms in $I4/m$ space group are $C_{4h}$ and $C_S$ (K), $S_4$ and $C_1$ (Fe), $C_4$ and $C_1$ (Se). Factor group analysis yields:
\begin{eqnarray}
(C_{4h}): \Gamma = A_u+E_u, \nonumber \\
(C_S): \Gamma = 2A_g+A_u +2B_g+B_u+E_g+2E_u, \nonumber \\
(S_4): \Gamma = A_u+B_g+E_g+E_u, \nonumber \\
(C_4): \Gamma = A_g+A_u+E_g+E_u, \nonumber \\
(C_1): \Gamma = 3A_g+3A_u+3B_g+3B_u+3E_g+3E_u. \nonumber
\end{eqnarray}
\noindent One can expect 27 Raman active phonons originating from the $I4/m$ phase: 9A$_g(\alpha_{x'x'+y'y'}, \alpha_{z'z'})$, 9B$_g(\alpha_{x'x'-y'y'}, \alpha_{x'y'})$ and 9$E_g(\alpha_{x'z}, \alpha_{y'z})$ modes. Due to $(001)$-orientation of our samples, only A$_g$ and B$_g$ modes can be observed in the Raman spectra.\cite{Sumpor} According to the analysis of Raman mode intensities,  A$_g$ phonons vanish in the crossed polarization configuration, whereas B$_g$ modes are observable in both crossed an parallel polarization configurations.\cite{Lazarevic1}

\section{Results and discussion}

\begin{figure}
\begin{center}
\includegraphics[width = 0.47\textwidth]{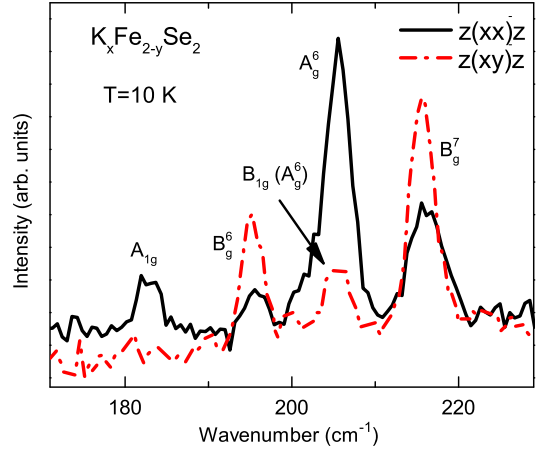}
\caption{(Color online) Raman scattering spectra of K$_x$Fe$_{2-y}$Se$_2$ single crystals measured at 10 K in parallel ($\mathbf{z(xx)\bar{z}}$) and crossed ($\mathbf{z(xy)\bar{z}}$) polarization configuration, in a frequency range from 170 to 230 cm$^{-1}$. Presence of the A$_g^6$ mode at about 206 cm$^{-1}$ prevents unambiguous assignation of the low-intensity Raman mode appearing at almost the same frequency in crossed polarization configuration.}
\label{fig3}
\end{center}
\end{figure}

Figure~\ref{fig2} (a) shows polarized Raman scattering spectra of K$_{x}$Fe$_{2-y}$Se$_2$  single crystals, in the spectral range between 150 and 325 cm$^{-1}$, measured from the (001)-plane of the sample at various temperatures. The observed Raman modes are assigned according to \cite{Lazarevic1} (see Figure~\ref{fig2} (a))  and in agreement with previously discussed selection rules for the high symmetry ($I4/mmm$) and low symmetry ($I4/m$) phase of this phase separated sample.\cite{Priroda} Raman active mode appearing in crossed polarization configuration at about 206 cm$^{-1}$ (marked by arrow in Figure~\ref{fig3}) could be assigned as the B$_{1g}$ symmetry mode.\cite{Lazarevic1} However, high-intensity A$_g^6$ phonon is present in parallel polarization configuration at almost the same frequency (see Figure~\ref{fig3}). Due to the possible leakage of the A$_g^6$ mode, unambiguous assignment of the marked Raman mode cannot be done here.

Polarized Raman scattering spectra of K$_{0.8}$Fe$_{1.8}$Co$_{0.2}$Se$_2$ single crystals, measured from the (001)-plane of the sample at various temperatures, are presented in Figure~\ref{fig2} (b). Energies of Raman active phonons are close to the energies of the corresponding modes in K$_x$Fe$_{2-y}$Se$_2$. Based on this, it can be concluded that doping of K$_x$Fe$_{2-y}$Se$_2$ single crystal with small amount of Co does not have significant impact on phonon spectrum in the normal state.

Ignatov \textit{et al.} \cite{Blumberg} argued that the Raman active mode at about 203 cm$^{-1}$ in K$_x$Fe$_{2-y}$Se$_2$ (at 3 K) disappears above 250 K, together with two other modes at about 163 and 210 cm$^{-1}$. They concluded that these modes belong to the $I4$ space group. As can be seen in Figure~\ref{fig2}, this phonon (206 cm$^{-1}$) is clearly visible in our Raman spectra of both compounds up to the room temperature, and it is assigned in \cite{Lazarevic1} as the B$_{1g}$ mode, originating from the $I4/mmm$ phase. Due to the absence of new Raman modes below 250 K, there are no indications of structural phase transition in our samples.

Behavior of the phonon modes with temperature can be properly described in terms of the phonon self-energy:\cite{Cardona2}
\begin{equation}
\Sigma_i (T) = \Delta_i(T)+i\Gamma_i(T),
\end{equation}
\noindent where $\Delta_i(T)$ represents change of the Raman mode energy and $\Gamma_i(T)$ is the mode linewidth, which is inversely proportional to the phonon lifetime $\tau$.

Therefore, temperature dependance of Raman mode energy can be described with:
\begin{equation}
\omega_i(T) = \omega_{0,i}+\Delta_i(T),
\end{equation}
\noindent where $\omega_{0,i}$ is temperature independent contribution to the energy of the phonon mode $i$ and $\Delta_i(T)$ can be decomposed as:\cite{Cardona2, Hackl}
\begin{equation}
\Delta_i(T) = \Delta_i^V+\Delta_i^A.
\end{equation}
The first term in (3) represents change of phonon energy due to the thermal expansion of the crystal lattice, and is given by\cite{Cardona2}
\begin{equation}
\Delta_i^V = \omega_{0,i} \big(e^{-3\gamma_i \int_{0}^{T} \alpha(T')dT'}-1\big),
\end{equation}
\noindent where $\gamma_i$ is the Gr\"{u}neisen parameter of a given mode and $\alpha(T)$ is volume thermal expansion coefficient of a material under investigation.

\begin{figure}
\begin{center}
\includegraphics[width = 0.45\textwidth]{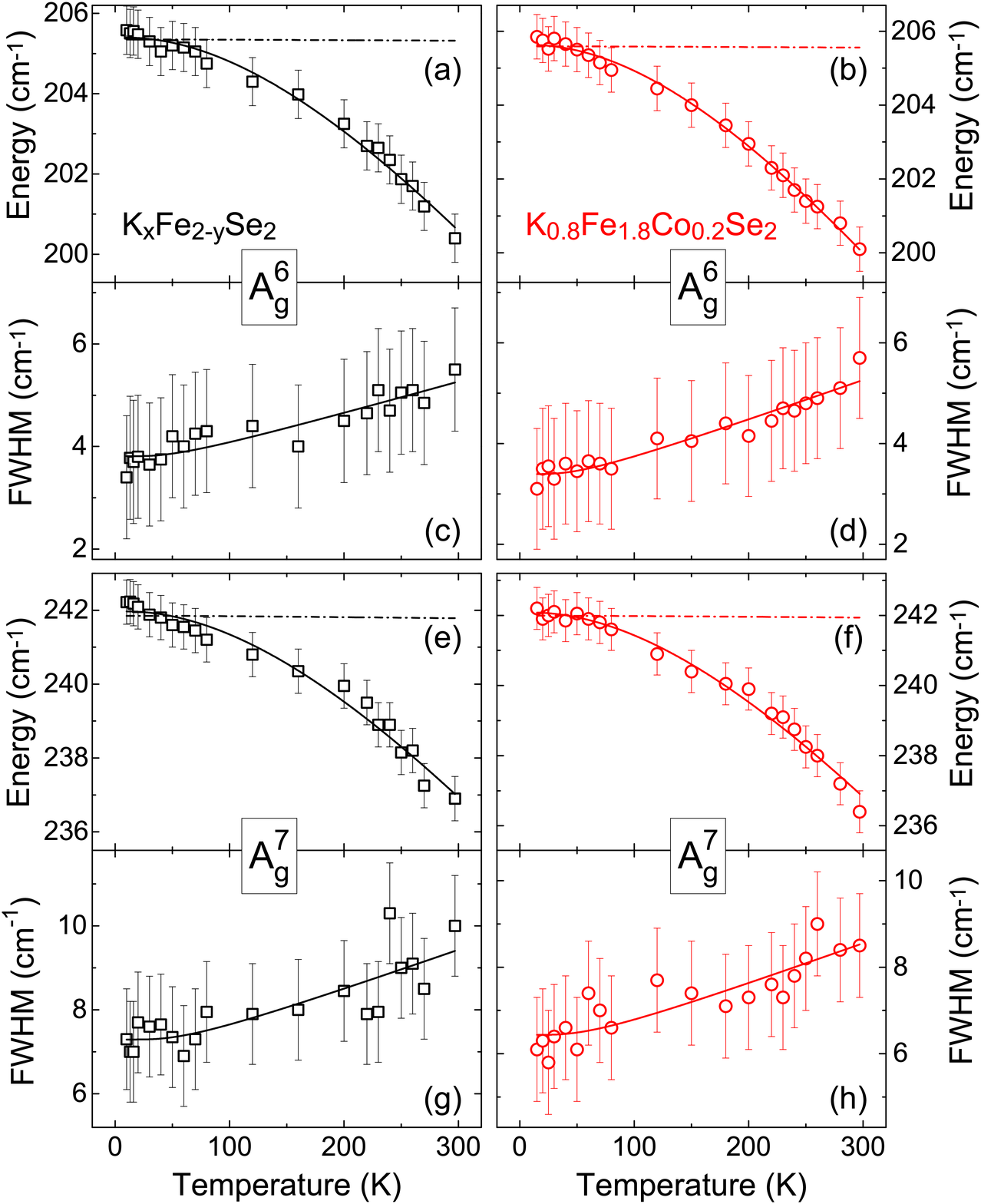}
\caption{(Color online) Energy and linewidth of A$_g^6$ and A$_g^7$ Raman modes of K$_x$Fe$_{2-y}$Se$_2$ (left panel) and K$_{0.8}$Fe$_{1.8}$Co$_{0.2}$Se$_2$ single crystals (right panel) as a function of temperature. Solid lines represent theoretical dependences, according to (2) and (6), with $\Delta^A$ neglected. A$_g^6$ and A$_g^7$ mode energy dependence on temperature with only anharmonic contribution included is shown by dash-dot lines.}
\label{fig4}
\end{center}
\end{figure}

The second term in (3) describes contribution of phonon-phonon scattering (lattice anharmonicity) to the Raman mode energy. If we describe anharmonic effects by three-phonon processes, it follows:\cite{Hackl}
\begin{equation}
\Delta_i^A = -C_i\big(1+\frac{4\lambda_{ph-ph,i}}{e^{\hbar \omega_0/2k_BT}-1}\big).
\end{equation}
\noindent $C$ and $\lambda_{ph-ph}$ are anharmonic constant and phonon-phonon interaction constant, respectively.

\begin{figure}
\begin{center}
\includegraphics[width = 0.48\textwidth]{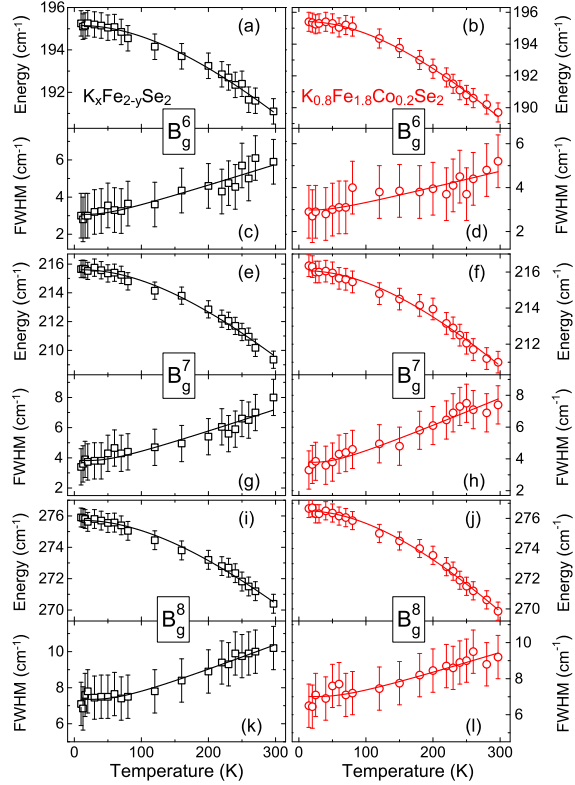}
\caption{(Color online)  Temperature dependence of energy and linewidth for B$_{g}^6$, B$_{g}^7$ and B$_{g}^8$ modes of K$_x$Fe$_{2-y}$Se$_2$ (left panel) and K$_{0.8}$Fe$_{1.8}$Co$_{0.2}$Se$_2$ single crystals (right panel). Solid lines are plotted according to (2) (taking into account only volume contribution to the phonon mode energy) and (6).}
\label{fig5}
\end{center}
\end{figure}

Temperature dependence of Raman mode linewidth is caused only by phonon anharmonicity:\cite{Hackl}
\begin{eqnarray}
\Gamma_i(T) = \Gamma_{0,i}\big(1+\frac{2\lambda_{ph-ph,i}}{e^{\hbar \omega_0/2k_BT}-1}\big),
\end{eqnarray}
\noindent where $\Gamma_0$ is the anharmonic constant.

\begin{table*}[t]
\caption{Best fit parameters for temperature dependence of energy and linewidth for the observed Raman modes ($\omega_0$ - temperature independent phonon energy,  $\gamma$ - Gr\"{u}neisen parameter, $\Gamma_0, C$ - anharmonic constants, $\lambda_{ph-ph}$ - phonon-phonon interaction constant). Temperature dependence of linewidth has not been analyzed for the A$_{1g}$ and B$_{1g}$ modes due to large relative errors.}
\label{tab1}
\centering
\resizebox{0.9\textwidth}{!}{%
\begin{tabular}{c c c c c c c c}\br
\multicolumn{8}{c}{K$_x$Fe$_{2-y}$Se$_2$} \\ \br

& A$_{1g}$ & A$_g^6$ & A$_g^7$ & B$_g^6$ & B$_{1g}$(A$_g^6$) & B$_g^7$ & B$_g^8$ \\ [1mm]

\cline{2-8} \\[-1.0em]

$\omega_0$ (cm$^{-1}$) & 182.2(2) & 205.43(6) & 242.06(9) & 195.15(5) & 205.34(9) & 215.62(7) & 275.79(6) \\ [1mm]

$\gamma$ & 1.6(1) & 1.74(5) & 1.52(6) & 1.57(4) & 1.63(8) & 2.10(5) & 1.42(4) \\ [1mm]

$\Gamma_0$ (cm$^{-1}$) & & 2.8(1) & 7.2(2) & 1.9(1) & & 3.2(1) & 6.31(6) \\ [1mm]

$\lambda_{ph-ph}$ & & 0.29(3) & 0.13(2) & 0.42(5) & & 0.36(3) & 0.25(2) \\ [1mm]

$C$ (cm$^{-1}$) & & 0.018(2) & 0.106(4) & 0.009(1) & & 0.023(2) & 0.072(2) \\ \br

\multicolumn{8}{c}{K$_{0.8}$Fe$_{1.8}$Co$_{0.2}$Se$_2$} \\ \br

& A$_{1g}$ & A$_g^6$ & A$_g^7$ & B$_g^6$ & B$_{1g}$(A$_g^6$) & B$_g^7$ & B$_g^8$ \\ [1mm]

\cline{2-8} \\[-1.0em]

$\omega_0$ (cm$^{-1}$) & 183.14(6) & 205.63(5) & 242.16(7) & 195.55(5) & 206.06(7) & 216.08(7) & 276.56(6) \\ [1mm]

$\gamma$ & 2.57(6) & 2.05(4) & 1.61(5) & 1.94(3) & 2.07(6) & 1.81(5) & 1.78(3) \\ [1mm]

$\Gamma_0$ (cm$^{-1}$) & & 2.31(7) & 6.3(2) & 1.40(9) & & 3.1(1) & 6.6(1) \\ [1mm]

$\lambda_{ph-ph}$ & & 0.28(3) & 0.15(3) & 0.61(7) & & 0.44(4) & 0.17(2) \\ [1mm]

$C$ (cm$^{-1}$) & & 0.0130(9) & 0.083(4) & 0.0052(7) & & 0.023(2) & 0.080(3) \\ \br

\end{tabular}}
\end{table*}

Parameter $C$ is connected with $\omega_0$ and $\Gamma_0$ via relation:\cite{Hackl}
\begin{equation}
C_i = \frac{\Gamma_{0,i}^2}{2\omega_{0,i}}.
\end{equation}
\noindent $\omega_0$ and $\Gamma_0$ can be determined by the extrapolation of the corresponding experimental data to 0 K. With these parameters known, we can fit the phonon mode linewidth, using (6), to obtain $\lambda_{ph-ph}$. Then, by determining parameter $C$ via (7), Raman mode energy can be properly fitted, with $\gamma$ as the only free parameter.

To the best of our knowledge, thermal expansion coefficient $\alpha(T)$ of K$_x$Fe$_{2-y}$Se$_2$ is unknown. Because of that, we used experimental data for FeSe, given in Figure~\ref{fig2} of \cite{FeSefit}.

Energy and linewidth temperature dependence of the A$_g^6$ and A$_g^7$ modes of K$_x$Fe$_{2-y}$Se$_2$ and K$_{0.8}$Fe$_{1.8}$Co$_{0.2}$Se$_2$ single crystals are presented in Figure~\ref{fig4}. Using (2), (4), (6) and (7), and following the previously described procedure, we have obtained best fit parameters, which are shown in Table~\ref{tab1}. It can be seen that parameter $C$  has very small values for both modes, thus contribution to the Raman mode energy from the phonon-phonon interaction (as can be seen in Figure~\ref{fig4} (a), (b), (e) and (f)) can be neglected. For this reason, change of Raman mode energy with temperature is properly described only with the thermal expansion term $\Delta^V$ (solid lines in Figure~\ref{fig4} (a), (b), (e) and (f)).

Figure~\ref{fig5} shows energy and linewidth of B$_g^6$, B$_g^7$ and B$_g^8$ phonon modes of K$_x$Fe$_{2-y}$Se$_2$ and K$_{0.8}$Fe$_{1.8}$Co$_{0.2}$Se$_2$ single crystals as a function of temperature. Analysis of temperature dependence of energy and linewidth for these modes was carried out as in the case of the A$_g$ phonons. Best fit parameters are listed in Table~\ref{tab1}. Small values of the parameter $C$ allow us to omit contribution from the lattice anharmonicity to the phonon mode energy. Good agreement of theoretical curves with experimental data justifies our assumption. Gr\"{u}neisen parameter is close to the conventional value $\gamma$$\sim$2 for B$_g$ modes, as well as for the other analyzed Raman active phonons.\

\begin{figure}
\begin{center}
\includegraphics[width = 0.47\textwidth]{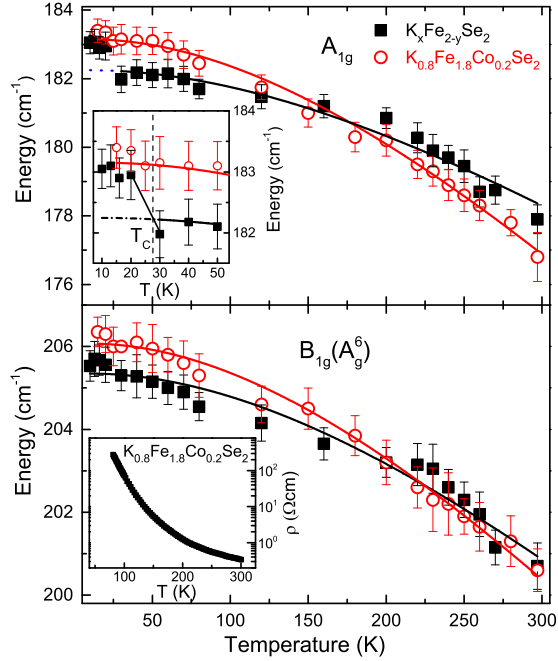}
\caption{(Color online) Temperature dependence of energy for A$_{1g}$ and B$_{1g}$(A$_g^6$) Raman modes of K$_x$Fe$_{2-y}$Se$_2$ and K$_{0.8}$Fe$_{1.8}$Co$_{0.2}$Se$_2$ single crystals. Upper inset: An enlarged view of A$_{1g}$ mode energy dependence on temperature in a low temperature region near $T_C$. Lower inset: Temperature dependence of the electrical resistivity for K$_{0.8}$Fe$_{1.8}$Co$_{0.2}$Se$_2$ shows nonmetalic behavior of this single crystal.}
\label{fig6}
\end{center}
\end{figure}

The dependence of A$_{1g}$ and B$_{1g}$(A$_g^6$) mode energy  of K$_x$Fe$_{2-y}$Se$_2$ and K$_{0.8}$Fe$_{1.8}$Co$_{0.2}$Se$_2$ single crystals on temperature is given in Figure~\ref{fig6}. Rather small values of intrinsic linewidth for both modes ($\Gamma_0$$\sim$2 cm$^{-1}$) give negligible values of parameter $C$. Therefore, A$_{1g}$ and B$_{1g}$(A$_g^6$) phonon energy temperature dependence is analyzed using (2) with only $\Delta^V$ term included. B$_{1g}$(A$_g^6$) mode energy of both single crystals exhibits a good agreement between experimental data and expected behavior (represented by solid lines in Figure~\ref{fig6}) in the whole temperature range. However, the A$_{1g}$ mode energy temperature dependence is well described with the proposed model only in the case of the Co- doped (non-superconducting) sample. The abrupt change in the A$_{1g}$ mode energy around $T_C$ in the superconducting sample (see inset of Figure~\ref{fig6}) have been observed. This results in a clear deviation from the expected behavior described with Eq. (2) and is governed by some other interaction.

It is well established that upon entering superconducting state, some of the phonons are renormalized due to the change in electronic structure with the superconducting gap opening, and therefore electron-phonon coupling. Depending on the values of the corresponding phonon energies and the superconducting gap magnitude, it may result in hardening or softening of the phonons.\cite{CardonaHTc} Recently, the superconductivity-induced self-energy effect have also been reported in some iron arsenides.\cite{Litvinchuk2, Kwon, Lemmens} We believe that the A$_{1g}$ mode hardening around $T_C$, observed for the superconducting and absent for the non-superconducting sample (see inset of Figure~\ref{fig6}), is caused by superconductivity induced phonon renormalizaton. Moreover, the hardening around $T_C$ has been observed only for the mode corresponding to the vibrations of the superconducting high symmetry ($I4/mmm$) phase in the superconducting K$_x$Fe$_{2-y}$Se$_2$ sample, and it is in agreement with the expected behavior ($2\Delta_c<\omega _{A_{1g}}$, $2\Delta_c$$\sim$130 cm$^{-1}$).\cite{Priroda,Gap,CardonaHTc}

From the observed renormalization of the A$_{1g}$ mode energy at $T_C$, the strength of electron-phonon coupling in K$_x$Fe$_{2-y}$Se$_2$ single crystal can be estimated in $\Gamma$-point. By using the formula $\lambda = -\kappa \textrm{Re}(\sin u/u)$\cite{Zeyher1, Lemmens}, where $\kappa = \frac{\omega^{SC}}{\omega^N}-1\approx 0.55\%$ is the superconductivity-induced renormalization constant ($\omega^{SC}$ and $\omega^N$ are phonon energies in superconducting and normal state, respectively), and $u = \pi+2i\cosh^{-1}\left(\frac{\omega^N}{2\Delta}\right)$, we obtained $\lambda_{A_{1g}}^{\Gamma}\approx 0.002$, indicating weak electron-phonon interaction.

As can be seen from Figures~\ref{fig4},~\ref{fig5} and~\ref{fig6}, connection between electronic transport channels in K$_x$Fe$_{2-y}$Se$_2$, which manifests as a broad hump in the in-plane resistivity $\rho_{ab}(T)$\cite{Superconductivity, Hall, Precipitate}, does not have a significant impact on the energy and linewidth temperature dependences for all investigated Raman active modes.

\section{Conclusion}
In this article Raman scattering study of superconducting K$_x$Fe$_{2-y}$Se$_2$ and nonsuperconducting K$_{0.8}$Fe$_{1.8}$Co$_{0.2}$Se$_2$ single crystals has been presented. Two Raman active modes from the superconducting ($I4/mmm$) phase
and seven phonon modes from the insulating ($I4/m$) phase are observed in the investigated frequency range. The same number of the observed modes in these two compounds, together with their similar energies, suggests that in K$_{0.8}$Fe$_{1.8}$Co$_{0.2}$Se$_2$ single crystal phase separation is also present. Temperature-induced change of Raman mode linewidth is in good agreement with the lattice anharmonicity model. Behavior of Raman mode energy as a function of temperature for all the analyzed modes is well described by the contribution from the lattice thermal expansion alone. Sudden change of the A$_{1g}$ mode energy near $T_C$ in K$_x$Fe$_{2-y}$Se$_2$ is due to the rearrangement of the electronic states to which this mode couples as superconducting gap opens. From the superconductivity-induced A$_{1g}$ phonon energy renormalization rather small electron-phonon coupling constant in $\Gamma$-point is obtained.

\ack

We gratefully acknowledge discussions with R. Hackl.
This work was supported by the Serbian Ministry of Education, Science and Technological Development under Projects ON171032 and III45018, as well as the Serbian-Germany bilateral project ''Competition between s-wave and d-wave pairing in FeSe''. Part of this work was carried out at the Brookhaven
National Laboratory which is operated for the Office of Basic Energy
Sciences, U.S. Department of Energy by Brookhaven Science Associates
(DE-Ac02-98CH10886) and by the Center for Emergent Superconductivity, an Energy Frontier Research Center funded by the U.S. DOE, Office for Basic Energy Science.

\bibliographystyle{unsrt}

\end{document}